\documentclass[prb,showpacs,floatfix,twocolumn]{revtex4}

\usepackage{graphicx}

\begin{document}

\title[Magnetodielectric phenomena in CoSeO$_4$]
{Magnetic ordering and magnetodielectric phenomena in CoSeO$_4$} 

\author{Brent C. Melot, Lucy E. Darago, and Ram Seshadri}
\affiliation{Materials Department and Materials Research Laboratory\\
	 University of California, Santa Barbara CA 93106}

\author{Abby Goldman}
\affiliation{Physics Department, Mount Holyoke College, \\
South Hadley, Massachusetts 01075}

\author{Joshua D. Furman}
\affiliation{Department of Materials Science and Metallurgy, \\ 
University of Cambridge, U.K. CB2 3QZ}

\author{Efrain E. Rodriguez}
\affiliation{National Institute of Standards and Technology, \\
Gaithersburg, Maryland 20899-6102}

\begin{abstract}

CoSeO$_4$ has a structure consisting of edge-sharing chains of 
Co$^{2+}$ octahedra which are held together by SeO$_4^{2-}$ tetrahedra 
{\it via} shared oxygen atoms at the edges of the octahedra.
DC magnetization measurements indicate a transition to an ordered state below 
30\,K.  Powder neutron diffraction refinements suggest 
an ordered state with two unique antiferrromagnetic chains within the unit 
cell. Isothermal magnetization measurements indicate a temperature-dependent 
field-induced magnetic transition below the ordering temperature. From 
neutron diffraction, we find this corresponds to a realignment of spins 
from the canted configuration towards the $c$-axis. The dielectric constant 
shows a change in slope at the magnetic ordering temperature as well as a quadratric 
dependence on the external magnetic field.
% indicating an 
%interplay between the spin and charge degrees of freedom.
% as well as a quadratric 
%dependence on the external magnetic field.

\end{abstract}


\maketitle

\section{Introduction}

Magnetodielectrics are materials in which the dielectric properties couple to 
changes in the magnetic order. The ability to find and design new single-phase
materials which exhibit this coupling between the spin and charge degrees of 
freedom has significant technological implications in the development of 
magnetic sensors and field-tunable dielectrics.~\cite{Lawes2009,Hur2004a} 
Systems showing incommensurate and noncollinear spin structures such as 
CoCr$_2$O$_4$~\cite{Yamasaki2006,Lawes2006}, Mn$_3$O$_4$~\cite{Tackett2007}, 
SeCuO$_3$~\cite{Lawes2003}, and Ni$_3$V$_2$O$_8$~\cite{Lawes2005} have all been extensively 
studied to understand the nature of such interactions. Recent work 
characterizing the magnetoelectric coupling in the magnetic chain compounds 
MnWO$_4$,\cite{Taniguchi2006} and LiCu$_2$O$_2$.\cite{Masuda2005,Park2007} 
indicate that systems with reduced crystallographic dimensionality should 
be of particular interest because of the strong interplay between the charge, 
lattice, and magnetic degrees of freedom that occurs in these 
materials.\cite{Cheong2007,Kimura2007} Reduced dimensionality has been found 
to result in noncollinear and canted antiferromagnetic spin structures which 
break symmetry restrictions and allow electric polarization to develop.

\begin{figure}[t]
\centering \includegraphics[height=7.5cm]{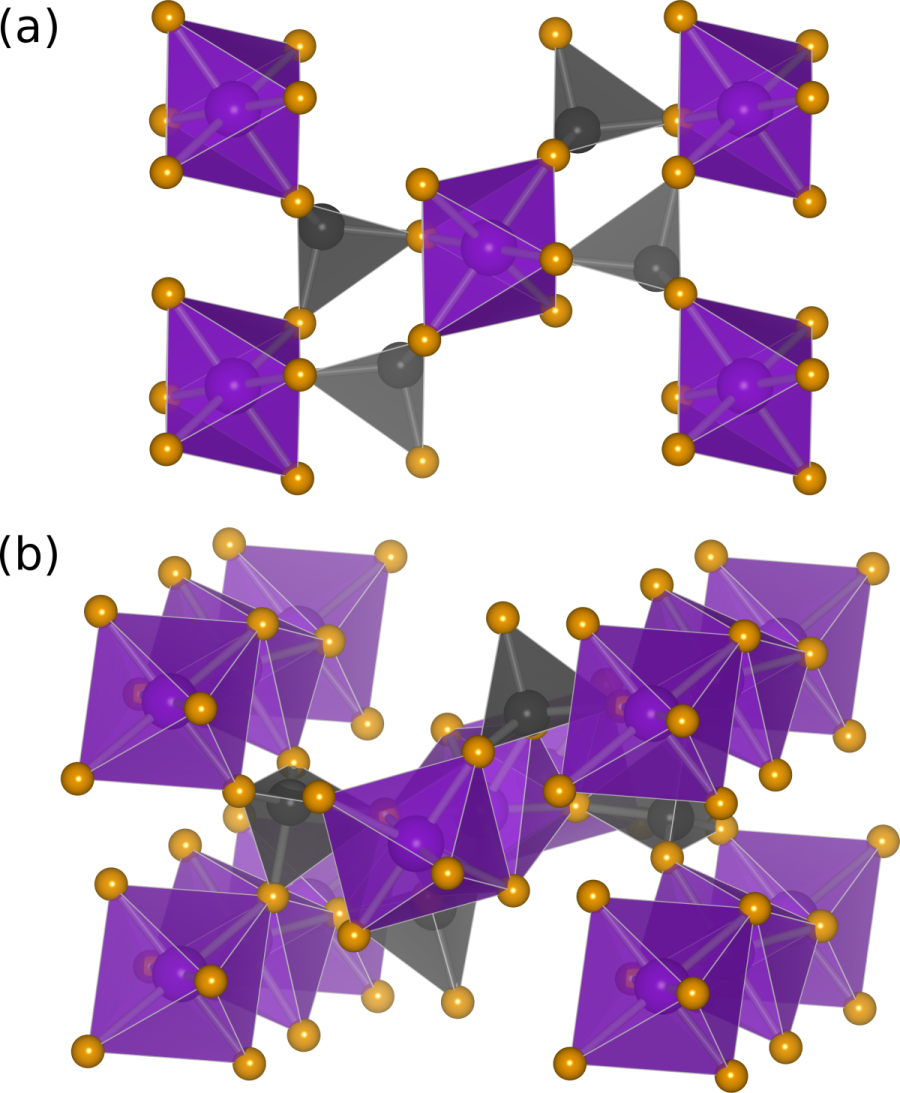}
\caption{(Color Online) Illustration of the crystal structure of CoSeO$_4$. 
Chains of octahedrally coordinated Co$^{2+}$ (purple) are bound together by 
SeO$_4$ tetrahedra {\it via} shared oxygen atoms at the edges of the octahedra.
Note that the chains are alternately tilt $+35^{\circ}$ and $-35^{\circ}$ off 
the $a$-axis. (a) View down the $c$-axis of the unit cell. (b) View down the 
$c$-axes to illustrate the chains of edge sharing CoO$_6$}
\label{fig:struct1}
\end{figure}

Here we discuss the preparation and detailed magnetic and dielectric 
characterization of CoSeO$_4$; a magnetic chain compound which adopts a 
crystal structure analagous to $\beta$-CoSO$_4$ (Fig.~\ref{fig:struct1}).
The orthorhombic unit cell contains chains of octahedrally coordinated 
Co$^{2+}$ which are bound together by SeO$_4^{2-}$ tetrahedra {\it via} 
shared oxygen atoms at the edges of the octahedra.
The chains of octahedra are tilted with respect to neighboring chains with the 
axis of the octahedra tilted $-$35$^{\circ}$ and $+$35$^{\circ}$ off the $a$-axis 
for the chains on the edge and in the center of the cell respectively.
Previously reports have characterized the nuclear and magnetic structure using 
neutron diffraction,\cite{Fuess1968,Fuess1969} but no reports on the magnetic 
susceptibility or dielectric properties have been made to date, and the 
effects of magnetic fields have not been fully investigated.
Using detailed magnetization, heat capacity, powder neutron diffraction, and 
dielectric measurements, we show that the transition to long-range 
antiferromagnetic below 30\,K is accompanied by a signficant change in the 
slope of the temperature dependence of the dielectric constant. We also find 
a field-induced change in the magnetic structure. 
The dependence of the dielectric constant on the external field, which is quadratic in nature,
changes dramatically above the field-induced magnetic transition.

\section{Experimental methods}

The title compound was prepared following a previously reported 
procedure\cite{Koleva1997} by neutralizing a solution of selenic acid 
(2.559\,g 23.06\,mmol) of SeO$_2$ dissolved in 50\,cm$^3$ of water) with 
Co$_2$CO$_3$(OH)$_2$ (1.629\,g, 7.69\,mmol) at 70$^{\circ}$C and 
recrystallizing in water.  The final crystallization step was performed 
by allowing the remaining water to evaportate at room temperature over several 
days. Faceted crystals of dark red CoSeO$_4\cdot$6H$_2$O up to 2\,cm on a side 
were obtained from this proceedure.
These crystals were ground and dehydrated at 125$^{\circ}$C overnight to yield 
a bright pink powder of CoSeO$_4\cdot$H$_2$O which was subsequently heated at 
315$^{\circ}$C for several days to produce a light violet powder of 
anhydrous CoSeO$_4$.
Co$_2$CO$_3$(OH)$_2$ was prepared following a separate previously reported 
procedure.~\cite{Tanaka1992} Stoichiometric amounts of CoNO$_3\cdot$5H$_2$O 
and Na$_2$CO$_3$ were separately dissolved in H$_2$O with a 2:1 
(NaCO$_3$:CoNO$_3$) volume excess. 
The solution of CoNO$_3$ was added to the solution of 
Na$_2$CO$_3$ which had been preheated to 90$^{\circ}$C.
After stirring for 2 hours the resulting dark purple precipitate was collected 
by vacuum filtration and subsequently washed with H$_2$O and ethanol.
Purity of the precursors and final product was confirmed by powder X-ray 
diffraction on a Philips XPERT MPD diffractometer operated at 45\,kV and 40\,mA.

Temperature dependence of the DC magnetization was measured on well-ground 
powder samples using a Quantum Design MPMS 5XL SQUID magnetometer. 
The specific heat data were collected using the semiadiabatic technique as 
implemented in a Quantum Design Physical Property Measurement System (PPMS), 
under zero applied field as well as under a 50\,kOe field. 
The measurement was made by mixing the compound with equal parts by mass of Ag 
powder and pressing into a pellet in order to improve thermal coupling to the 
stage.  The contribution from Ag was measured separately and subtracted.
Variable temperature neutron diffraction data under a magnetic field were 
collected on the BT1 powder diffractometer at the National Institute of 
Standards and Technology, Gaithersburg, MD using a wavelength of 2.08\,\AA\/.
Dielectric properties were measured by attaching polished copper electrodes to 
opposite faces of a cold-pressed pellet of CoSeO$_4$, using a Quantum Design 
PPMS for temperature and field control, and an Agilent 4980A LCR meter to 
measure the capacitance.

\begin{figure}[t]
\centering \includegraphics[width=7cm]{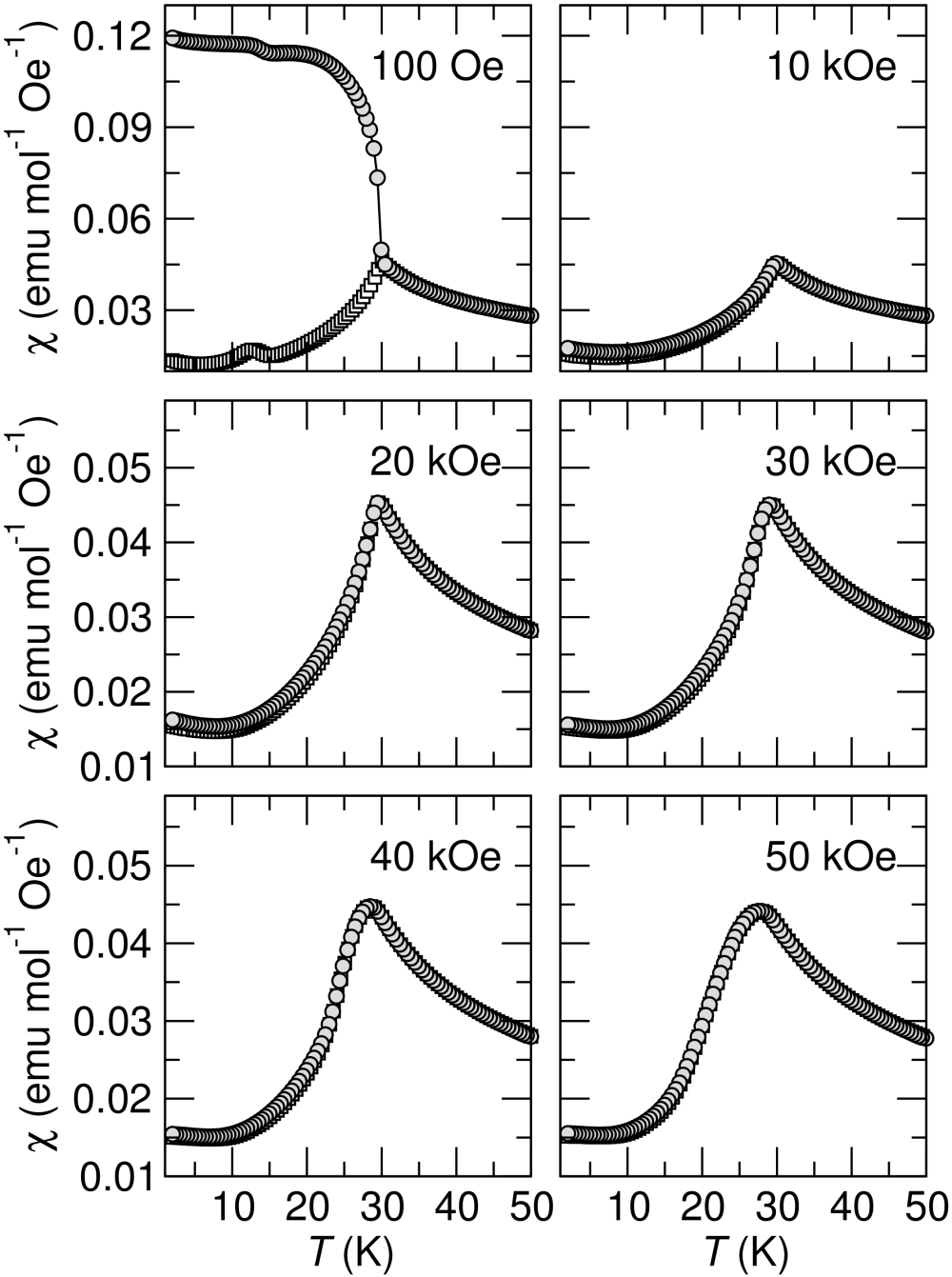}
\caption{Field-cooled (FC, open symbols) and zero-field cooled (ZFC,
closed symbols) magnetic susceptibility of a powder sample of CoSeO$_4$
acquired under increasing strengths of the external magnetic field.
At 30\, K, the system orders to an antiferromagnetic state. A weak
ferromagnetic component is found in the 100\,Oe data, suggested from the
separating ZFC and FC traces. Note also that the cusp which indicates the
magnetic ordering is smeared out with increasing field strength.}
\label{fig:fc}
\end{figure}

\section{Computational methods}

Density functional theory (DFT) calculations were performed using the Vienna {\it Ab-initio}
Simulation Package (VASP)~\cite{Kresse1996,Kresse1999} at the experimental lattice parameters. 
The projector augmented-wave (PAW) method~\cite{Blochl1994} was used together with the local density approximation
(LDA) Ceperley-Alder exchange correlation functional.~\cite{Ceperley1980}
A plane-wave energy cutoff of 500\,eV and a 6$\times$3$\times$9 $\Gamma$-centered Monkhorst-Pack\cite{Monkhorst1976} 
$k$-point mesh corresponding to 162 irreducible $k$-points was used to sample 
the Brillouin zone. The tetrahedron method with the Bl\"ochl 
correction\cite{Blochl1994} was used for Brillioun zone integrations.
Correlation was treated using the LDA+$U$ formalism within the rotationally 
invariant approach of Liechtenstein \textit{et al.}\cite{Liechtenstein1995}
The applicability of the LDA+$U$ formalism is validated in the example of $d^7$
Co$^{2+}$ because crystal symmetry results in an absence of 
orbital degeneracy. Several values of $U$ were tested with only qualitative 
changes to the band structure being found. Therefore a $U$ of 5\,eV and a $J$ 
of 1\,eV was chosen to agree with previously published DFT results calculated 
on octahedrally coordinated Co$^{2+}$.~\cite{Ederer2006}

\section{Results and Discussion} 

The high temperature region (200\,K to 300\,K) of the inverse susceptibility 
was fit to the Curie-Weiss equation, $C/(T-\Theta_{CW})$.
A Curie-Weiss temperature of $-$36\,K and an effective moment $\mu_{eff}$ of 
4.38\,$\mu_B$ was determined for the data collected under a 10\,kOe field.
The effective moment lies between the spin-only value of 3.87\,$\mu_B$ and 
the value of 5.2\,$\mu_B$ expected for octahedrally coordinated Co$^{2+}$ 
($d^7$, $t_{2g}^{5}e_g^2$, $\mathbf{S}$=3/2, $\mathbf{L}$=3) with a fully 
unquenched orbital contribution as obtained using the relationship
$\mu_{L+S} = \sqrt{4\mathbf{S}(\mathbf{S}+1) + \mathbf{L}(\mathbf{L}+1)}$.\cite{Day1960} 
The existence of an orbital contribution is common from high-spin $d^7$ 
systems in an octahedral $d^7$ crystal field due to the orbital degeneracy 
in the $t_{2g}$ levels. The small reduction of the orbital contribution likely 
arises from the irregular octahedral environment of the magnetic sites reducing 
some hybridization with the surrounding oxygen anions.

\begin{figure}[t]
\centering \includegraphics[width=8cm]{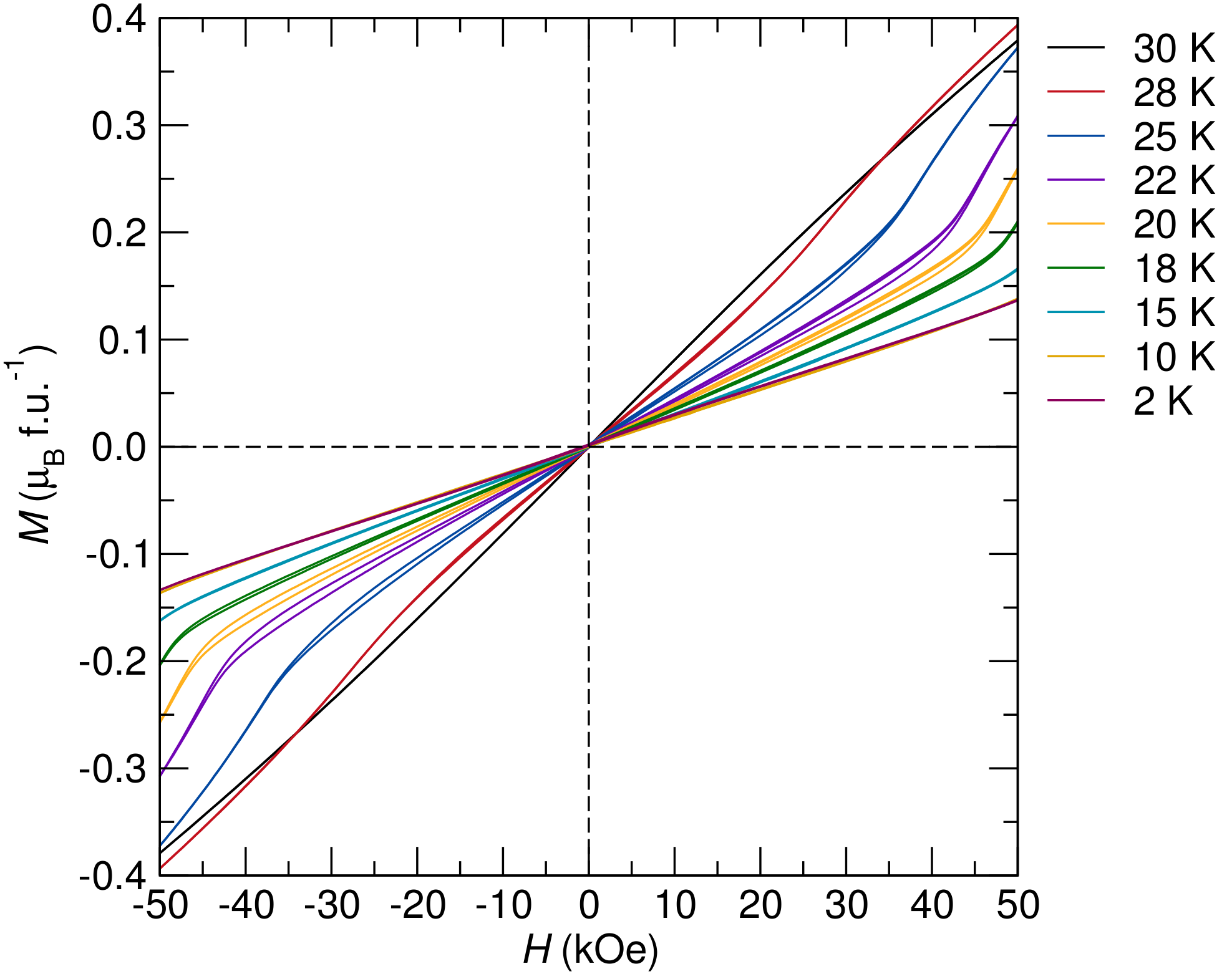}
\caption{(Color Online) (a) Isothermal magnetization loops at various 
temperatures below the N\'eel temperature. Note the field induced transition 
which pushes to higher fields at low temperatures.}
\label{fig:hys}
\end{figure}

Fig.\,\ref{fig:fc} shows the temperature dependence of the magnetic 
susceptibility of CoSeO$_4$ under a variety of magnetic field strengths.
The spins begin to order below 30\,K which is very near the expected 
Curie-Weiss temperature indicating that there is very little frustration from 
competing exchange interactions.  When measured in a 100\,Oe field, a sudden 
jump in the susceptibility occurs at 30\,K which can be attributed to some 
weak ferromagnetism which results from a canted antiferromagnetic arrangement 
of the spins. Measuring the sample in larger fields minimizes the contribution 
from the weak-ferromagnetic component and gives rise to the cusp-like behavior 
expected from a well-compensated antiferromagnet. This cusp is found to 
broaden with increasing field strength indicating that the magnetic order may 
be driven to a different ground state in the presence of sufficiently large 
external fields. To further examine the nature of the field-dependence of the 
magnetic order, isothermal magnetization loops were collected at a variety of 
temperatures (Fig.\,\ref{fig:hys}). An upturn from the linear field-dependence 
expected for a well-behaved antiferromagnet is found beginning in a field of 
approximately 20\,kOe at 28\,K. The upturn shows a strong temperature 
dependence, with the field required to induce it increasing to 45\,kOe by 
18\,K and subsequently out of the 50\,kOe range of the magnetometer used here
by 10\,K.  Such a field-induced magnetic transition may be attributed to 
overcoming of the magnetocrystalline anisotropy of the octahedral Co$^{2+}$ 
cations which would manifest as a spin-flop transition where the spins realign 
from one preferred axis to another.

\begin{figure}[t]
\centering \includegraphics[width=8cm]{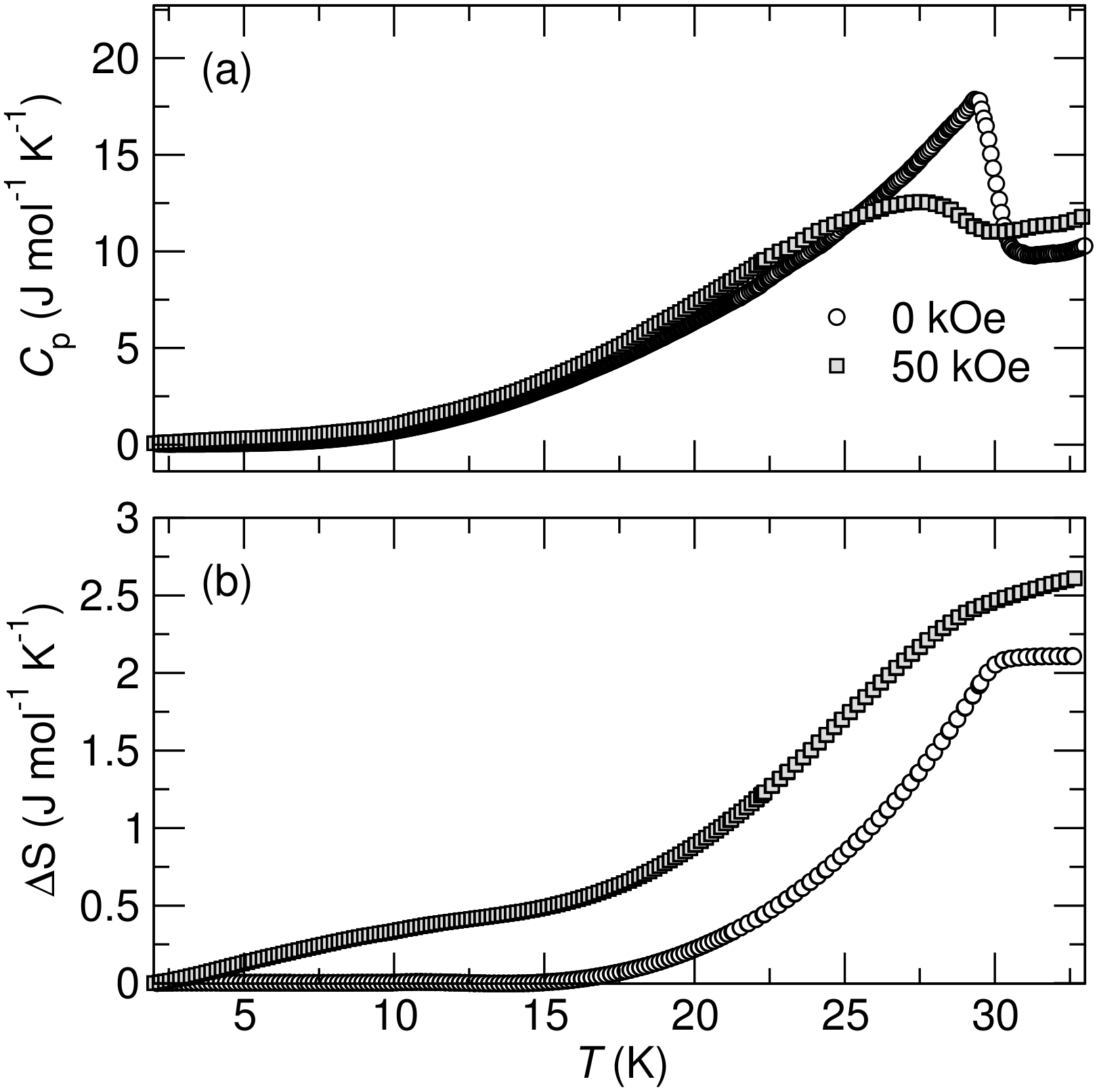}
\caption{(a) Temperature dependence of the specific heat of CoSeO$_4$
measured on a powdered sample in $H$ = 0\,kOe and 50\,kOe field. (b) Entropy 
released due to the magnetic ordering, obtained by integration of the magnetic 
heat capacity.}
\label{fig:hc}
\end{figure}

Fig.\,\ref{fig:hc} (a) shows the specific heat of CoSeO$_4$ measured in zero 
field, and under a an external magnetic field of 50\,kOe. The specific heat 
shows a sharp anomaly at the magnetic ordering temperature of 30\,K in the 
absence of an external magnetic field. Application of a 50\,kOe field results 
in a smearing out of the transition in a manner similar to the behavior 
observed in the susceptibility data shown in Fig.\,\ref{fig:fc} 
The lattice contribution to the specific heat was approximated by fitting the 
data above the magnetic transition to a polynomial expansion 
($\beta$T$^3$+$\gamma$T$^5$+$\delta$T$^7$). Subtracting the lattice 
contribution and integrating $C_{p}/T$ yields a change in entropy due to 
magnetic ordering of 2.1\,J\,mol$^{-1}$\,K$^{-1}$ and 
2.5\,J\,mol$^{-1}$\,K$^{-1}$ in zero field and a 50 kOe field respectively.  
This value is significantly smaller than the value of 
11.5\,J\,mol$^{-1}$\,K$^{-1}$ predicted by the Boltzmann equation 
($\Delta S = R\ln(2\mathbf{S}+1)$, $\mathbf{S}=3/2$).
This considerable diminution of the spin entropy from what is expected
may indicate that the reduced dimensionality of the system prevents 
complete ordering of the spins, or that possibly the order within chains is 
not well correlated with that in neighbording chains.

\begin{figure}[t]
\centering \includegraphics[width=9cm]{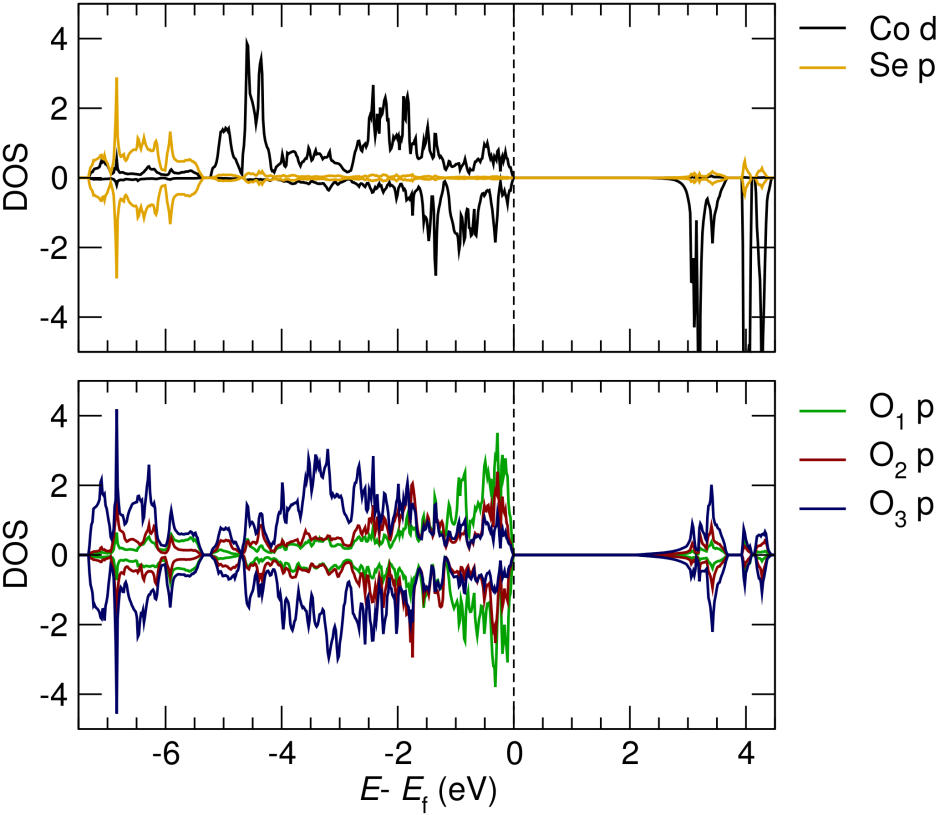}
\caption{(Color Online) Densities of state of CoSeO$_4$ from a collinear spin structure 
calculation. The magnetic structure used in the calculation was that of 
antiferromagnetic chains aligned antiparallel to neighboring chains.}
\label{fig:DOS}
\end{figure}

To better understand the electronic structure and chemical bonding in the 
title compound, density functional theory calculations were performed using 
the LDA$+U$ formalism. Fig.\,\ref{fig:DOS} illustrates the density of states 
(DOS) calculated for the experimental room temperature structure of CoSeO$_4$.
The DOS show a significant degree of overlap between the Se-$p$ and O-$p$
states and between Co-$d$ and O-$p$ states below the Fermi energy. However, 
there is a gap of approximately 250\,meV between the Co and Se states.
It can be seen that a majority of the Se and O overlap is related to bonding 
between Se and O$_3$. The O3 position forms two bonds within the Se 
tetrahedron which are 1.56\,\AA; short in comparison with the Se--O1 and 
Se--O2 bond lengths which are 1.74\,\AA\/ and 1.78\,\AA\/ respectively.  
Such bonding is consistent with the idea that the SeO$_4$ tetrahedra contains 
two double and two single bonds with surrounding oxygens to satisfy its octet.
It is interesting to note that the Se-O3 bonds correspond to the shared oxygens
at the corners of the CoO$_6$ octahedra within a single chain.
This arrangement of double bonds can be expected to significantly influence 
magnetic superexchange between the CoO$_6$ chains. 

\begin{figure}[t]
\centering \includegraphics[height=8cm]{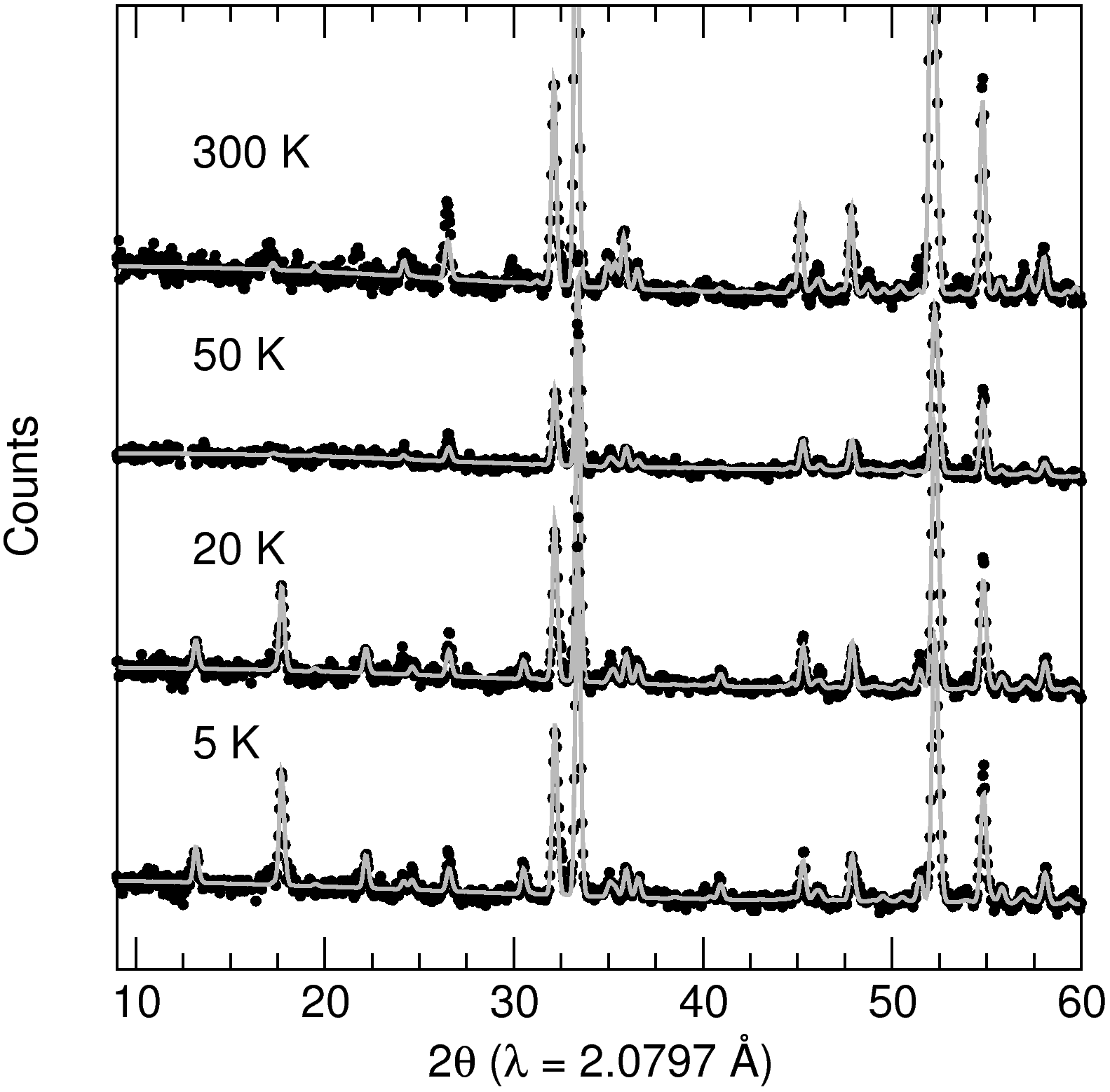}
\caption{Low angle region of the powder neutron-diffraction patterns of
CoSeO$_4$ (BT1,NIST) obtained at different temperatures.}
\label{fig:neutron}
\end{figure}

The DOS calculated using the LDA formalism results in metallic densities of state, 
but by accounting for correlations within LDA+$U$, with $U$ = 5\,eV and $J$ = 1\,eV, 
a gap of approximately 3\,eV opens between the occupied and unoccupied minority spins.
We also note that the band gap which is approximately 2\,eV corresponds to the gap 
between the highest filled states and unoccupied O-$p$ orbitals.    
The distorted octahedral environment of the Co$^{2+}$ breaks the 
degeneracy of the unoccupied $e_{g}$ orbitals corresponding to $d_{z^2}$ and 
$d_{x^2-y^2}$ with unoccupied $d_{xz}$ above the Fermi energy.

\begin{figure}[b]
\centering \includegraphics[height=7cm]{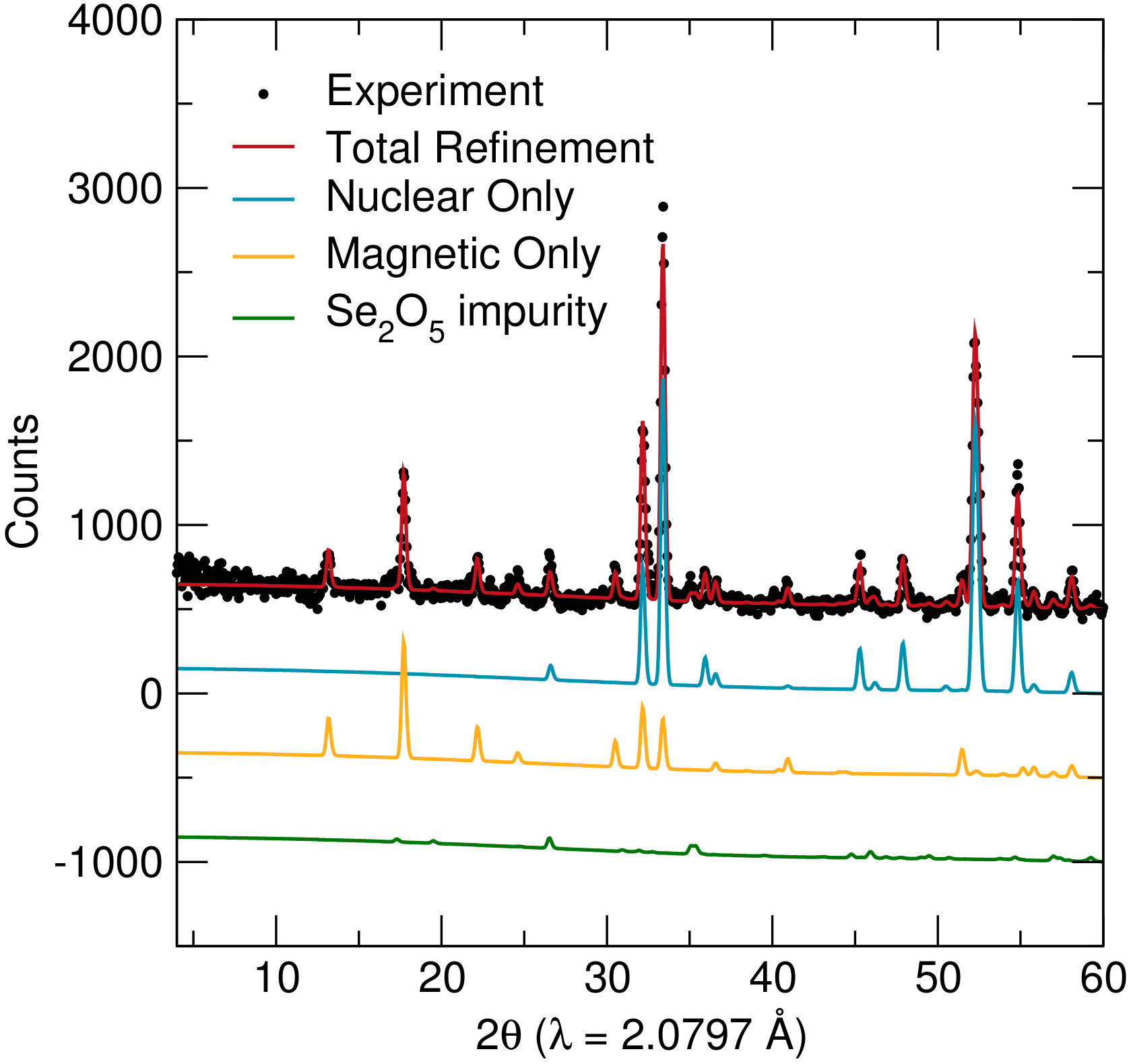}
\caption{(Color Online) Low angle region of the powder neutron-diffraction pattern of
CoSeO$_4$ obtained at 5\,K and broken down in to the contribution from the nuclear, magnetic, and
a small non-magnetic impurity phase of Se$_2$O$_5$.}
\label{fig:neutron2}
\end{figure}

The antiferromagnetic state is stabilized by more than 0.5\,eV compared to a ferromagnetic alignment of spins.
The calculated spin magnetic moment on each Co is 2.72\,$\mu_B$ which is slightly 
reduced from the full value of 3\,$\mu_B$ expected for a $d^7$ cation in an octahedral 
coordination environment.  
Such a reduction may be associated with the covalent bonding to the surrounding oxygens
or may be related to the fact that the calculation is restricted to a collinear spin 
configuration which is not the true ground state observed {\it via } neutron diffraction.

\begin{table}[t]
\caption{Summary of the results of Rietveld structure refinement of variable
temperature neutron diffraction data obtained in zero magnetic field.}
\label{tab:cell}
\begin{tabular}{rlll}
\hline \hline
                        & 300\,K     &  20\,K     &   5\,K  \\
\hline 
$a$ (\AA)               & 9.0421(3)  & 9.0355(4)  & 9.0344(4)  \\
$b$ (\AA)               & 6.7669(3)  & 6.7397(3)  & 6.7398(3)  \\
$c$ (\AA)               & 4.8860(2)  & 4.8759(2)  & 4.8757(2)  \\
$M$ ($\mu_B$)           &    --      & 3.14(8)    & 3.50(8) \\
$V$ (\AA$^3$)           & 298.964(2) & 296.927(2) & 296.886(2) \\
$R_{\mbox{nuc.}}$ (\%)  & 8.3        & 8.8        & 8.4 \\
$R_{\mbox{mag.}}$ (\%)  &    --      & 15.7       & 14.1 \\
\hline \hline
\end{tabular}
\end{table}

\begin{figure}[b]
\centering \includegraphics[height=10cm]{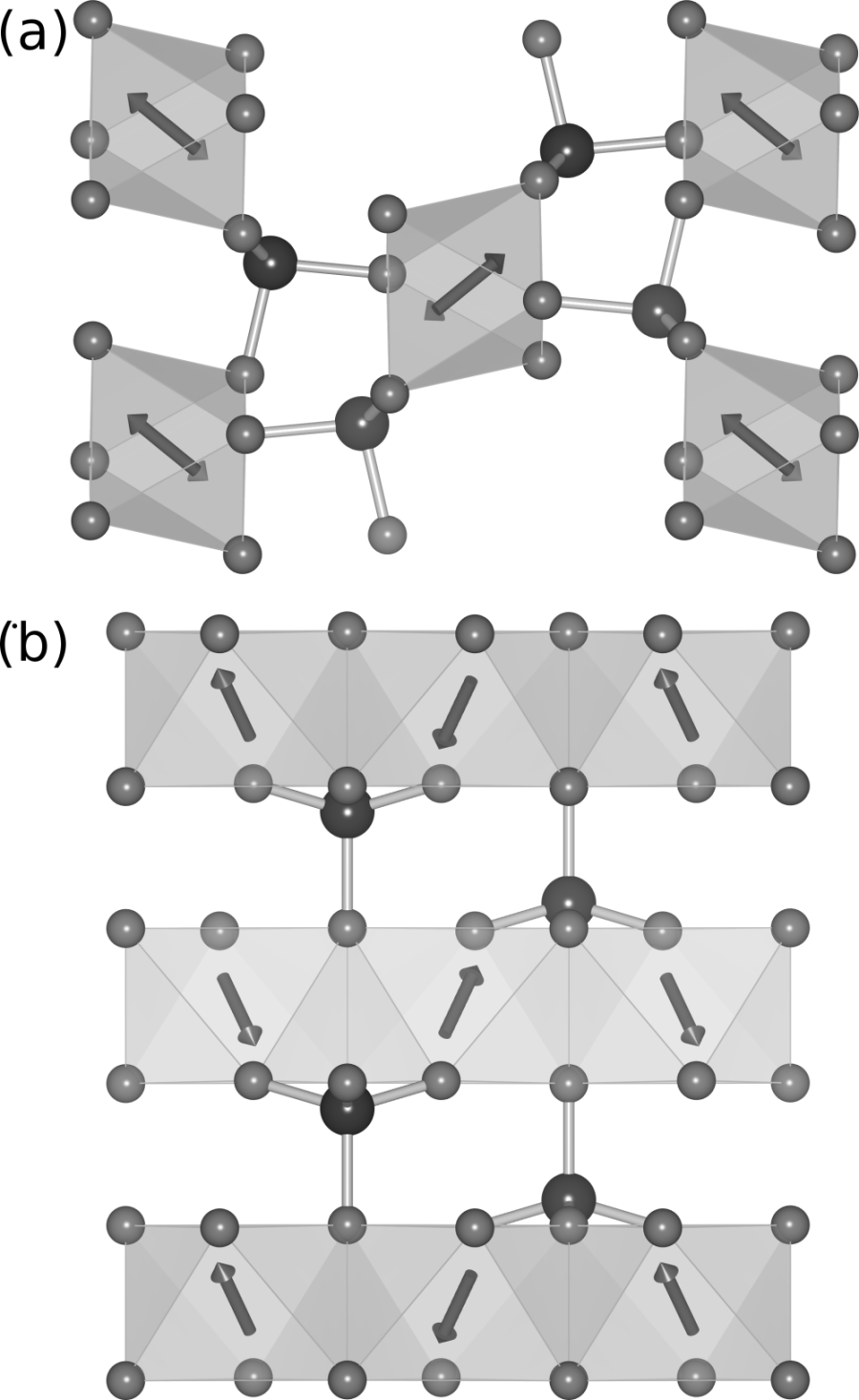}
\caption{Proposed magnetic structure of CoSeO$_4$ as determined from Rietveld
refinements of the neutron diffraction pattern obtained using a
$\lambda$ = 2.08\,\AA\/ at 2\,K. (a) View down the $b$-axis of the $Pnma$
structure. (b) View down the $c$-axis. The small light grey atoms are oxygen
while the larger and darker grey atoms are selenium}
\label{fig:magstruct}
\end{figure}

The thermal evolution of the neutron-diffraction patterns collected from 
300\,K to 5\,K are shown in Fig.\,\ref{fig:neutron} with 
Table\,\ref{tab:cell} showing a summary of refined parameters.
Three magnetic reflections appear below 10\,K at 12$^\circ$, 18$^\circ$, and 
23$^\circ$, all of which are consistent with a propagation vector 
$\mathbf{k}$ = 0. It was only possible to simultaneously fit all three 
magnetic peaks by using the basis vectors of the irreducible representation 
$\Gamma_1$. The resulting magnetic structure is illustrated in 
Fig.\,\ref{fig:magstruct}. The spins on each Co align antiferromagnetically 
down the length of the chain and with respect to the neighboring chains.
The moment of each spin has components along all three axes of the unit cell 
as listed in Table \ref{tab:magn} with a total magnetic moment of 
3.59\,$\mu_B$ at 5\,K. 
The magnetic structure in the absence of a magnetic field shown in 
Fig.\,\ref{fig:magstruct} agrees well with the previously reported structure of Fuess.~\cite{Fuess1969} 
We note that while Fuess reported the structure in $Pbnm$ we have chosen the standard
setting of $Pnma$.  
The resulting spin configuration (compared to that reported by Fuess) has spins oriented largely in the $ac$ 
plane with the moments tilting 60$^\circ$ (60$^\circ$) off the $a$-axis, 70$^\circ$ (69$^\circ$) off the $b$-axis, and 44$^\circ$ (41$^\circ$) off the $c$-axis again at 5\,K. The moment within individual chains do not cancel 
completely with a resulting uncompensated moment along the $a$-axis which is 
canceled by the neighboring chains. This uncompensated moment within the 
individual chains may be a source of the weak ferromagnetism observed in small 
external magnetic fields. 

The effect of applying a magnetic field of 70\,kOe is to first reduce the 
total magnetic moment to 2.60\,$\mu_B$ at 5\,K.
The field also causes the spins to reorient with respect to the unit cell.
At 20\,K the moments form angles of 66$^\circ$ off the $a$-axis, 64$^\circ$ off the $b$-axis, and 
37$^\circ$ off the $c$-axis. From examination of the isothermal magnetization 
measurements in conjunction with the reorientated moments the field-induced magnetic transition appears
to corresponds to a spin-flop transition on to the $c$-axis.  

\begin{table}[t]
\caption{Components of the magnetic moment on each Co in the magnetic 
structure described by $\Gamma_1$.  Note that Co at the origin occupies
a 4$a$ Wyckoff site. The Co within the unit cell are: Co1 (0, 0, 0), 
Co2 ($\frac{1}{2}$, 0, $\frac{1}{2}$), Co3 (0, $\frac{1}{2}$, 0), 
Co4 ($\frac{1}{2}$, $\frac{1}{2}$, $\frac{1}{2})$} 
\label{tab:magn}
\begin{tabular}{rl|rrr|rrr}
\hline \hline
       &      &        & 0\,kOe  &        &        & 70\,kOe&        \\ 
\hline
       &      &  $M_x$ &  $M_y$  & $M_z$  &  $M_x$ & $M_y$  &  $M_z$ \\
\hline
20\,K  & Co1  & 1.51   &  1.05   & 2.51   &  0.86  & 0.93   &  1.68  \\
       & Co2  & $-$1.51  &  $-$1.05  & 2.51   &  $-$0.86 & $-$0.93  &  1.68  \\
       & Co3  & $-$1.51  &  1.05   & $-$2.51  &  $-$0.86 & 0.93   &  $-$1.68 \\
       & Co4  & 1.51   &  $-$1.05  & $-$2.51  &  0.86  & $-$0.93  &  $-$1.68 \\
\hline
5\,K   & Co1  & 1.76   &  1.19   & 2.90   &  1.24  & 1.14   &  1.98  \\
       & Co2  & $-$1.76  &  $-$1.19  & 2.90   &  $-$1.24 & $-$1.14  &  1.98  \\
       & Co3  & $-$1.76  &  1.19   & $-$2.90  &  $-$1.24 & 1.14   &  $-$1.98 \\
       & Co4  & 1.76   &  $-$1.19  & $-$2.90  &  1.24  & $-$1.14  &  $-$1.98 \\
\hline \hline
\end{tabular}
\end{table}

\begin{figure}[b]
\centering \includegraphics[width=7cm]{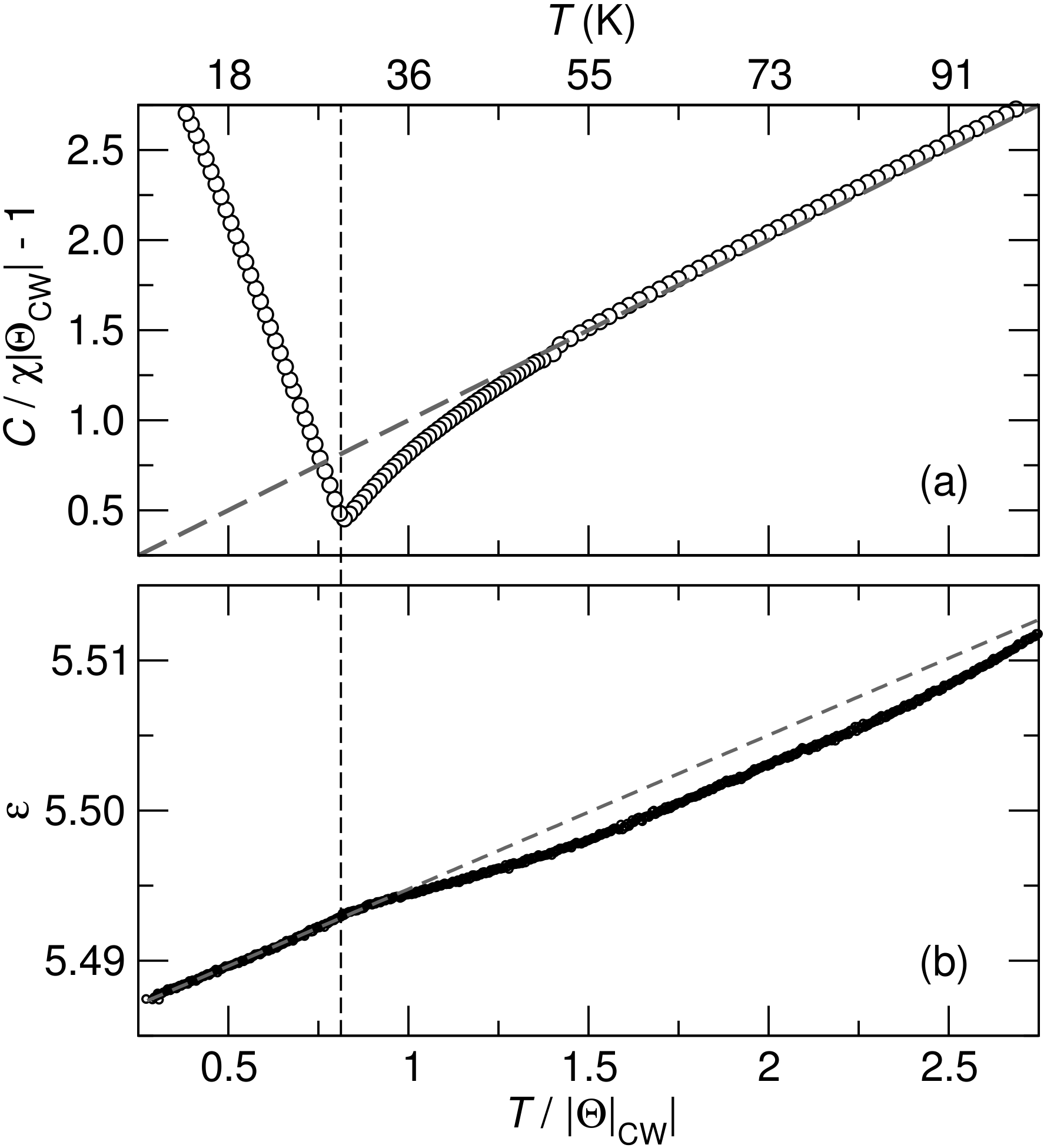}
\caption{(a) Inverse magnetic susceptibility normalized by the values extracted 
from fits to the Curie-Weiss formula in the high temperature region. Note 
that this manner of plotting emphasizes short-range correlations which cause 
deviations from ideal Curie-Weiss behavior which is illustrated as the dashed 
line. (b) Temperature dependence of the dielectric constant of CoSeO$_4$
measured on a pressed pellet of a polycrystalline sample at a frequency of 
1\,MHz. The dashed line is a guide-to-the-eye to emphasize the transition 
at the magnetic ordering temperature.} 
\label{fig:curie-eps}
\end{figure}

Fig.\,\ref{fig:curie-eps} (a) shows the inverse magnetic susceptibility 
normalized by the fits of the high temperature region (between 200\,K and 
300\,K) to the Curie-Weiss formula. Normalizing in this manner emphasizes 
deviations from the Curie-Weiss equation in the form of short-range 
correlations between the spins which develop above the ordering 
temperature.\cite{Melot2009a} Fig.\,\ref{fig:curie-eps}(b) shows the 
temperature dependence of the dielectric constant with a very clear 
change in slope in the dielectric constant at the magnetic ordering 
temperature. 

\begin{figure}[t]
\centering \includegraphics[width=7cm]{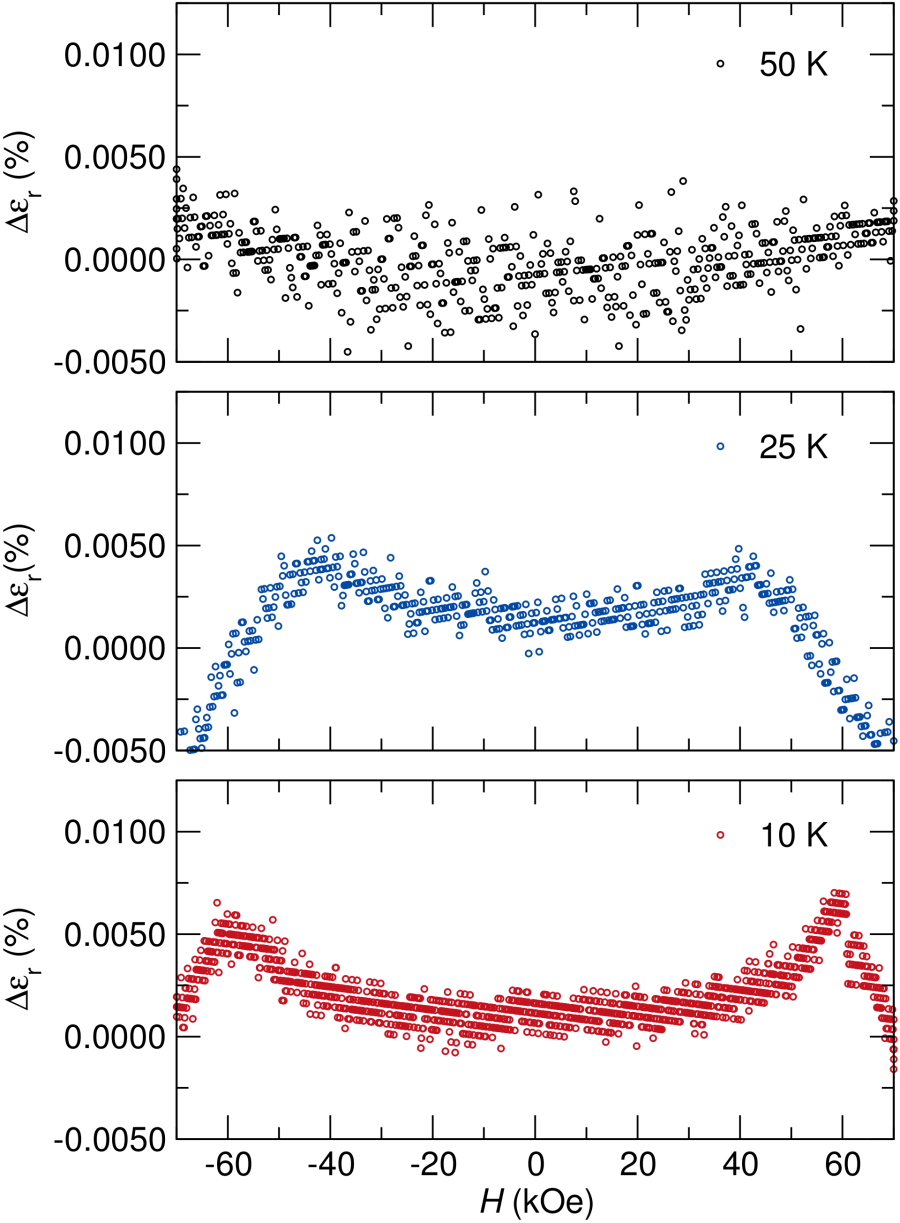}
\caption{Change in the dielectric constant measured at 1\,MHz plotted as 
a function of the strength of the external magnetic field at different temperatures.
Note the quadratic dependence until the field-induced magnetic transition }
\label{fig:dielectric}
\end{figure}

To further characterize the coupling between 
the magnetism and dielectric properties, the dielectric constant was measured 
as a function of external magnetic field. A quadratic dependence of the 
dielectric constant with small fields was found and is shown in 
Fig.\,\ref{fig:dielectric}. Above the field-induced magnetic transition,
the dielectric constant no longer exhibits a quadratic increase but rather 
changes abruptly to what appears to be a linear decrease.
This could reflect the possibility that the
low-field magnetic structure allows for a magnetodielectric response     
whereas the high-field magnetic structure does not.
A similar temperature and field-dependent behavior has been reported by 
N\'{e}nert and coworkers~\cite{Nenert2008} for a hybrid Cr(II) inorganic-organic 
material. They also find a quadratic change with
external magnetic field above the magnetic ordering temperature as well as below
and attribute the behavior to the always allowed $P^2H^2$ order parameter.
In that aspect, our results differ in that we see no response in the dielectric constant 
in the paramagnetic regime.  

\section*{Summary}

We have discussed the preparation and detailed magnetic and dielectric 
characterization of a magnetic chain compound.
Magnetization and specific heat measurements show a transition to a 
long-range canted antiferromagnetic state below 30\,K and a 
temperature-dependent field-induced magnetic transition. 
Powder neutron diffraction shows the spins lay predominantly in the 
$ac$-plane with the moments tilting 60$^\circ$ off the $a$-axis, 70$^\circ$ 
off the $b$-axis, and 44$^\circ$ off the $c$-axis again at 5\,K.
Temperature and field dependent measurements of the dielectric constant show 
an anomaly at the magnetic ordering temperature, which, at fixed temperature,
is quadratic in the external magnetic field.
Temperature dependent measurements of the dielectric constant show 
an anomaly at the magnetic ordering temperature.
We also find that the field-induced magnetic transition causes an abrupt 
change in the field dependence of the dielectric constant.
%Such behavior likely indicates the changes in the dielectric constant are 
%indicative of a linear magnetoelectric coupling which is not compatible with 
%the high-field magnetic structure.

\section*{Acknowledgements} 

The authors thank Daniel P. Shoemaker and Jonathan Suen for assistance in dieletric measurements.
Support for this work came from the National Science Foundation through a
MRSEC award (DMR 0520415; support for BM and facilities), a Career Award 
(DMR 0449354) and the RISE program at the UCSB MRL (support of internships
for LD and AG). 

\bibliography{melot}

\end{document}